\documentclass[3p,times,procedia]{elsarticle}
\usepackage{nupha_ecrc}

\usepackage{bm}
\usepackage{amsmath, color}

\newcommand\diag{\operatorname{diag}}

\newcommand\p{\partial}

\newcommand\T{{\rm T}}

\newcommand\lm{{\lambda} }

\volume{00}

\firstpage{1}

\journalname{Nuclear Physics A}

\runauth{}

\jid{nupha}

\jnltitlelogo{Nuclear Physics A}

\usepackage{amssymb}

\usepackage[figuresright]{rotating}

\begin{document}

\begin{frontmatter}


 \title{}

\dochead{}

\title{
Charge-dependent correlations \\
from event-by-event anomalous hydrodynamics
}

 \author[address1]{Yuji Hirono}
 \author[address3]{Tetsufumi Hirano}
 \author[address1,address2]{Dmitri E. Kharzeev}

\address[address1]{
 Department of Physics and Astronomy, Stony Brook University,
 Stony Brook, New York 11794-3800, USA
}
\address[address3]{
 Department of Physics, Sophia University, Tokyo 102-8554, Japan
}
\address[address2]{
 Department of Physics and RIKEN-BNL Research Center,
 Brookhaven National Laboratory, Upton, New York 11973-5000, USA
}


\address{}

\begin{abstract}
 We report on our recent attempt of quantitative
 modeling of the Chiral Magnetic Effect (CME) in heavy-ion collisions.
 We perform 3+1 dimensional anomalous hydrodynamic simulations on an
 event-by-event basis, with 
 constitutive equations that contain the anomaly-induced effects. We
 also develop a model of the initial condition for the axial charge
 density that 
 captures the statistical nature of random chirality imbalances created
 by the color flux tubes. Basing on the event-by-event hydrodynamic
 simulations for hundreds of thousands of collisions, we calculate the
 correlation functions that are measured in experiments, and discuss how
 the anomalous transport affects these observables.
\end{abstract}


\end{frontmatter}


\section{Introduction}
\label{sec:intro}

The Chiral Magnetic Effect (CME)
\cite{Kharzeev:2004ey,Kharzeev:2007tn,Kharzeev:2007jp,Fukushima:2008xe}
has received considerable attention in
recent years, particularly in the context of heavy-ion collisions. The
anomaly-induced transport effects like the CME are macroscopic and are
incorporated into hydrodynamic equations giving rise to ``anomalous
hydrodynamics''  \cite{Son:2009tf}.
Theoretically, the CME is expected to occur in
heavy-ion collision experiments.
The data reported by STAR \cite{Abelev:2009ad, Abelev:2009ac} and PHENIX
\cite{ajitanand2010p}
collaborations at RHIC and ALICE collaboration \cite{Abelev:2012pa} at
the LHC show a
behavior consistent with the CME, but the quantitative understanding is
still lacking. In order to reach a definitive conclusion, a reliable
theoretical tool that can describe the charge-dependent observables is
indispensable.

In this work \cite{Hirono:2014oda}, we quantitatively evaluate the
observables to
detect the anomalous transport, basing on event-by-event simulations of
anomalous hydrodynamics. 
The observable of interest in this talk is a charge-dependent two-particle correlation
\cite{Voloshin:2004vk}, 
\begin{equation}
 \gamma_{\alpha \beta}
  = 
 \left<
  \cos
  \left(
   \phi^1_\alpha
   +
   \phi^2_\beta
   - 2 \Psi_{\rm RP}
  \right)	
 \right>, 
\end{equation}
where
$\phi^{i}_{\alpha}$ is the azimuthal angle of $i-$th particle ($i=1,2$)
with charge $\alpha \in \{+, -\}$, and 
$\Psi_{\rm RP}$ is the reaction plane angle for $v_2$.
Physical meaning of this observable is evident if we decompose
$\gamma_{\alpha \beta}$ as 
\begin{equation}
 \gamma_{\alpha \beta}
  = 
 \left<
  \cos
  \left(
   \phi^1_\alpha - \Psi_{\rm RP}
  \right)	
  \cos
  \left(
   \phi^2_\beta - \Psi_{\rm RP}
  \right)	
  \right>
  -
 \left<
  \sin
  \left(
   \phi^1_\alpha - \Psi_{\rm RP}
  \right)	
  \sin
  \left(
   \phi^2_\beta - \Psi_{\rm RP}
  \right)	
	 \right>
	 \equiv
  \left<v_1^\alpha v_1^\beta
	  \right>
	  - 
  \left< a_1^\alpha a_1^\beta \right>, 
\end{equation}
where $v_1^\alpha$ ($a_1^\alpha$) is the directed flow which is
parallel (perpendicular) to $\Psi_{\rm RP}$, respectively. 

Let us see how $a_1$'s behave in the presence of anomalous effects. 
In off-central collisions, the magnetic fields perpendicular to
$\Psi_{\rm RP}$ (on average) are created. If the CME occurs, a current should be
generated along the magnetic field, which would result in finite 
$a_1^+$
and 
$a_1^-$.
The direction of the current depends on the sign of the initial axial
charge, which is basically random, so the signs of $a_1$s are also
random. 
However, the signs of $a_1^+$ and $a_1^-$ tend to be opposite. 
Thus, the CME expectations are the following:
(1) $\left< a_1^+ \right> = \left< a_1^- \right> = 0$, because the sign
of initial axial charge is random;
(2) $\left< \left(a_1^\alpha \right)^2 \right>$ becomes larger in the
presence of the CME currents;
(3) $\left< a_1^+ a_1^- \right> < 0$, which indicates the anti-correlation between $a_1^+$ and $a_1^-$.

\section{Event-by-event anomalous hydrodynamic model for heavy-ion collisions}

The model consists of three parts: (i) anomalous-hydro evolution, (ii)
hadronization via Cooper-Frye formula, and (iii) calculation of the
observables. 
For the hydro part, we solve the equations of motion for a
dissipationless 
anomalous fluid,
$
\p_\mu T^{\mu\nu} = e F^{\nu \lm} j_{\lm}, \ 
\p_\mu j^\mu = 0, \
\p_\mu j_5^\mu = -C E_\mu B^\mu, 
$
where 
$
C \equiv \frac{N_c} {2 \pi^2} \sum_{f} q_f^2
$ is the anomaly constant, 
$E^\mu \equiv F^{\mu\nu} u_{\nu}$, $B^\mu \equiv \tilde F^{\mu\nu} u_{\nu}$
with $ \tilde F^{\mu\nu} = \frac{1}{2} \epsilon^{\mu\nu\alpha\beta}
F_{\alpha\beta}$.
The energy-momentum tensor and currents are written as 
$
 T^{\mu\nu} = (\varepsilon +p)u^\mu u^\nu - p \eta^{\mu\nu}, \
 j^\mu = n u^\mu + \kappa_B B^\mu,  \
 j_5^\mu = n_5 u^\mu + \xi_B B^\mu,
$
where $\varepsilon$ is the energy density, $p$ is the hydrodynamic
pressure, $n$ and $n_5$ are electric and axial charge densities, 
$e \kappa_B \equiv  C \mu_5 [ 1 - \mu n / (\varepsilon+p)]$ and 
$e \xi_B \equiv  C \mu [ 1 - \mu_5 n_5 / (\varepsilon +p)]$ are transport
coefficients for chiral magnetic/separation effects (CME/CSE), 
and $\eta^{\mu\nu} \equiv \diag \{1,-1,-1,-1\}$ is the Minkowski
metric.
In this work, the electromagnetic fields are not dynamical and
treated as background fields. 
As for the equation of state (EOS), 
we use that of an ideal gas of quarks and gluons. 

Let us specify the electromagnetic field configurations used to get the
results shown later. 
We take $B_y$ to be ($x$-axis is chosen to be the reaction plane angle
$\Psi_{\rm RP}$)
\begin{equation}
 e B_y( \tau, \eta_s ,  \bm x_\perp) = e B_0 \frac{b}{2R} \exp \left(
- \frac{x^2}{\sigma_x^2}
- \frac{y^2}{\sigma_y^2}
- \frac{\eta_s^2}{\sigma_{\eta_s}^2}
-\frac{\tau}{\tau_{\rm B}}
\right) , 
\label{eq:magnetic}
\end{equation}
where $\sigma_x$, $\sigma_y$, and $\sigma_{\eta_s}$ are the widths of
the field in $x$, $y$, and $\eta_s$ (space-time rapidity) directions,
$\tau_B$ is the duration time of the magnetic field, $R = 6.38 \ {\rm
fm}$ is the radius of a gold nucleus, and $b$ is the impact parameter. 
Other elements of $\bm B$ and $\bm E$ are set to zero. 
The widths are taken
so that the fields are applied only in the region where matter exists
as 
$\sigma_{x} = 0.8 \left(R - \frac{b}{2}\right)$, 
$\sigma_{y} = 0.8 \sqrt{R^2 - (b/2)^2 }$,
and $\sigma_\eta = \sqrt{2}$.
We set other parameters as 
$\tau_B = 3 \ {\rm fm}$ and $eB_0 = 0.5 {\rm GeV}^2$ 
in following calculations, 
which is equivalent to $ e B_y(\tau_{\rm in}, 0, \bm 0) \sim (3 m_\pi)^2$. 

By solving the hydrodynamic equations, we obtain a particle
distribution via the Cooper-Frye formula with freezeout temperature
$T_{\rm fo} = 160 \ {\rm MeV}$.
We produce the hadrons by the Monte-Carlo sampling based on that
distribution. 
Thus, one random initial condition results in the particles in an event.
We repeat this procedure many times and store the data of many events,
that are later used to calculate the charge-dependent correlation
functions.
We calculate fluctuations of $v_1$ and $a_1$ separately, with the
following expressions, 
\begin{equation}
\left<
 \left(v^\alpha_1\right)^2
\right> 
\equiv
\left<
\frac{1}{_M {\rm P}_2} 
\sum_{<i,j>}
\cos( \phi_i^\alpha - \Psi_{\rm RP} )
\cos( \phi_j^\alpha - \Psi_{\rm RP} )
	 \right>, \quad
\left<
\left(a^\alpha_1\right)^2
\right>
\equiv
\left<
\frac{1}{_M {\rm P}_2} 
\sum_{<i,j>}
\sin( \phi_i^\alpha - \Psi_{\rm RP} )
\sin( \phi_j^\alpha - \Psi_{\rm RP} )
\right>. 
\end{equation}
for the same-charge correlation, 
where $M$ is the number of
produced particles, $ _M P_2 = M(M-1)$, $\sum_{<i,j>}$ indicates the sum
over all the pairs,
and outer bracket means averaging over events. 
Similar expression is used for the opposite-charge correlation. 

%
%

It is an important issue to estimate the amount of axial charges at the
beginning of hydro evolutions.
The major sources of the initial chiralities are color flux tubes in
heavy-ion collisions.
When two nuclei collide, numerous color flux tubes are spanned between
them.
The anomaly equation, $\partial_\mu j^\mu_5 = C \bm E^a \cdot \bm B^a$,
determines the rate of the axial charge generation, 
so the rate is determined by the value of $\bm E^a \cdot \bm B^a$.
There is no preferred sign of $\bm E^a \cdot \bm B^a$ and 
it can be positive or negative for different color flux tubes. 

In order to incorporate this feature, we have made an extension to the
so-called MC-Glauber model.
For each binary collision, we assign $\pm 1$ randomly.
Each sign indicates to the sign of color $\bm E^a \cdot \bm B^a$ of
the flux tube. 
Then, we initialize the axial chemical potential as 
\begin{equation}
\mu_5 (\bm x_\T, \eta_{\rm s}) = 
C_{\mu_5} f(\eta_{\rm s}) 
\sum_{j=1}^{N_{\rm coll}(\bm x_\T)} 
X_j \  , 
\end{equation}
where $X_j$ are the signs of $\bm E^a \cdot \bm B^a$ randomly assigned to binary
collisions, 
and $C_{\mu_5}$ is a constant which expresses the typical
strength of the $\mu_5$,
and $
 f(\eta_{\rm s}) = \exp \left[
- \theta(|\eta_{\rm s}| - \Delta \eta_{\rm s}) 
\frac{
(|\eta_{\rm s}| - \Delta \eta_{\rm s})^2 }{\sigma^2_{\eta}}
\right] 
$. 
The sum is taken over the binary collisions happening on that point in
the transverse plane.
The strength of the color fields are of the order of the saturation
scale. Taking this into account, we chose 
$C_{\rm \mu_5} = 0.1 \ {\rm GeV}$ \cite{Hirono:2014oda}.

\begin{figure}[htbp]
\begin{center}
\includegraphics[width=60mm]{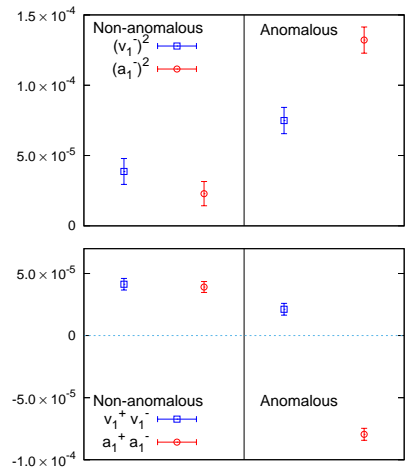}
\end{center}
 \caption{
 The correlations 
$\langle \left(v^{-}_1\right)^2 \rangle$, $\langle
 \left(a^-_1\right)^2 \rangle$ (upper figure), 
$\langle v_1^+ v_1^- \rangle$, and $\langle a_1^+ a_1^- \rangle$ (lower
 figure) 
for anomalous and non-anomalous cases 
 at $b=7.2 \ {\rm fm}$.
Those quantities are calculated from the data of $10,000$ events
for both of the anomalous and non-anomalous cases. 
}
\label{fig:vapm2}
\end{figure}

\section{Calculated observables}

The values of the observables are shown in Fig.~\ref{fig:vapm2}. 
The data from $10,000$ events are used to calculate those observables
for each of anomalous and non-anomalous case. Impact parameter is set to
$7.2 \ {\rm fm}$. 
The upper figure of Fig.~\ref{fig:vapm2} shows the values of
$\langle \left(v^{-}_1\right)^2 \rangle$ and
$\langle \left(a^-_1\right)^2 \rangle$.
In the left figure, anomalous transport effects are switched off (no
CME and CSE). 
The plots in the right figure are from anomalous hydrodynamic
simulations.
In the non-anomalous case, the values of the fluctuations of $v_1$ and $a_1$ are similar. 
When we switch on the anomaly (right figure), 
$\langle \left(v^{-}_1\right)^2 \rangle$ goes up,
and 
$\langle \left(a^-_1\right)^2 \rangle$ increases further.
The large fluctuation of $a_1$ is in line with the qualitative
expectation from the CME.
The order of magnitude of 
$\gamma_{\alpha \alpha}
=
\left<\left(v_1^\alpha\right)^2\right>
- \left<\left(a_1^\alpha\right)^2\right>
$ is comparable to experimentally measured values. 

In the lower figure of Fig.~~\ref{fig:vapm2}, we show the values of 
$\langle v_1^+ v_1^- \rangle$, and $\langle a_1^+ a_1^- \rangle$. 
In the absence of anomaly, they take similar positive values, but once we turn on
the anomaly, $\langle a_1^+ a_1^- \rangle$ becomes negative.
This is the indication of the anti-correlation between
$a^+_1$ and $a^-_1$ and is consistent with the CME expectations. 

%
It has been discussed that the observed values of
$\gamma_{\alpha\beta}$ might be reproduced by 
other effects unrelated to the CME,
including transverse momentum conservation
\cite{Bzdak:2010fd,Pratt:2010gy}, charge 
conservation \cite{Schlichting:2010na}, 
or cluster particle correlations \cite{Wang:2009kd}. 
Such effects are absent in the calculations here, because 
the particles are sampled based on the Cooper-Frye formula, which is
one-particle distribution, whereas all of the background effects arise from multi-particle correlations.
Thus, the difference between anomalous and non-anomalous calculations
purely originates from the CME and CSE. 
The contribution from the transverse momentum conservation
in the CME signal is recently estimated in Ref.~\cite{Yin:2015fca}, 
in which the charge deformations are treated as linear perturbations on
the bulk evolutions in 2+1D.

%

\section{Conclusions and outlook}

We reported the results of event-by-event simulations of
an anomalous hydrodynamic model for heavy-ion collisions. 
We solved the hydrodynamic equations including anomalous transport
effects (CME and CSE) in 3+1D, and calculated the values of observables.
We also developed a model of the initial axial charges created from the
color fux tubes. 
The caluculated values of the observables indicate that this observable
works as expected, and the order of magnitude is comparable to
experimentally measured values. 

The largest uncertainty arises from the choice of the life-time of the
magnetic fields.
The existence of conducting matter affects the duration of the
magnetic fields. 
We thus have to solve the hydrodynamic equations together with the Maxwell
equations -- this work is deferred to the future.

\section*{Acknowledgements}

This work was supported in part by the U.S. Department of Energy under Contracts No.
DE-FG-88ER40388 and DE-AC02-98CH10886. 
Y.~H. is supported by JSPS Research Fellowships for Young Scientists. 
The work of T.~H. was supported by JSPS KAKENHI Grants No. 25400269.


%

%
%






\bibliographystyle{elsarticle-num}
\bibliography{refs.bib}







\end{document}